\documentclass[twocolumn,showpacs,preprintnumbers,amsmath,amssymb]{revtex4}

\usepackage{graphicx}
\usepackage{dcolumn}
\usepackage{bm}
\usepackage{epsfig}

\begin{document}
\renewcommand{\thefootnote}{\fnsymbol{footnote}}
\renewcommand{\theequation}{\arabic{section}.\arabic{equation}}

\title{A closer look at arrested spinodal
decomposition in protein solutions}
\author{Thomas Gibaud$^{1}$
\footnote[1]{current address: Ecole Normale Sup\'{e}rieure de Lyon, 46 all\'{e}e d'Italie, 69364 Lyon cedex 07, France},
and Peter Schurtenberger$^{2}$}
 \affiliation{(1)Department of Physics, University of Fribourg, CH-1700
Fribourg, Switzerland.}
 \affiliation{(2)Adolphe Merkle Institute and Fribourg Center for Nanomaterials, University of Fribourg, CH-1700
Fribourg, Switzerland}

\begin{abstract}

Concentrated aqueous solutions of the protein lysozyme undergo a liquid-solid transition upon a temperature quench into the unstable spinodal region below a characteristic arrest temperature of $T_{f}$ = 15 $^{\circ}$C. We use video microscopy and ultra small angle light scattering in order to investigate the arrested structures as a function of initial concentration, quench temperature and rate of the temperature quench. We find that the solid-like samples show all the features of a bicontinuous network that is formed through a spinodal decomposition process. We determine the correlation length $\xi$ and demonstrate that $\xi$ exhibits a temperature dependence that closely follows critical scaling expected  for density fluctuations during the early stages of spinodal decomposition. These findings are in agreement with an arrest scenario based on a state diagram where the arrest or gel line extends far into the unstable region below the spinodal line. Arrest then occurs when during the early stage of spinodal decomposition the volume fraction $\phi_{2}$ of the dense phase intersects the dynamical arrest threshold $\phi_{2,Glass}$, upon which phase separation gets pinned into a space spanning gel network with a characteristic length $\xi$.

\end{abstract}
\pacs{pacs} \maketitle
\section{Introduction}

Gelation of colloidal particles has continued to attract considerable attention during the last few years. While it is ubiquitous and used in numerous applications  in material science, food science and the industry of personal care products \cite{2005natmat.Mezzenga}, it still remains a quite controversial topic and only partially understood. Under the general head line of "Dynamical arrest" or "Jamming", attempts have been made to arrive at a much more fundamental understanding of various fluid-solid transitions in colloidal suspensions that encompass  seemingly unrelated phenomena such as the glass transition observed for hard spheres, and gelation in suspensions of strongly attractive and thus irreversibly aggregating particles. Particular attention has been given to particles interacting via a short-ranged attractive potential, and the role of the key parameters volume fraction $\phi$, interparticle interaction strength $U_a$ and range of the potential $\Delta$ in determining the state diagram and in particular the gelation line has been addressed \cite{2002COCIS.Dawson, 2004COCIS.Trappe, 2007jpcm.Zaccarelli}.

Colloids interacting via short-ranged attractions exhibit a number of interesting phenomena such as a re-entrant glass transition at high values of $\phi$ and and a competition between phase separation and dynamical arrest at intermediate values of $U_a/kT$ and $\phi$ that can lead to the phenomenon of an arrested spinodal decomposition resulting in the formation of a solild-like network \cite{2007PRL.cardinaux, 2002nat.pham, 2008nat.lu, 2005PRL.Manley, 2007PRL.Buzzaccaro, 2004PRL.Kroy, 2003Lang.Bergenholtz, 1997physA.Verhaegh}. Moreover, a recent study of Lu et al. provided evidence that gelation for short range attractive particles would occur as a consequence of an initial equilibrium liquid-liquid phase separation that would then drive the formation of a space-spanning cluster that dynamically arrest to form a gel \cite{2008nat.lu}. However, while the phenomenon of an arrested spinodal decomposition as a possible mechanism for gelation in attractive colloidal suspensions appears to be generally accepted, there exists a controversy as to the location of the arrest line \cite{2007PRL.cardinaux, 2008nat.lu}.

\begin{figure}[]
\begin{center}
\includegraphics[width=240pt]{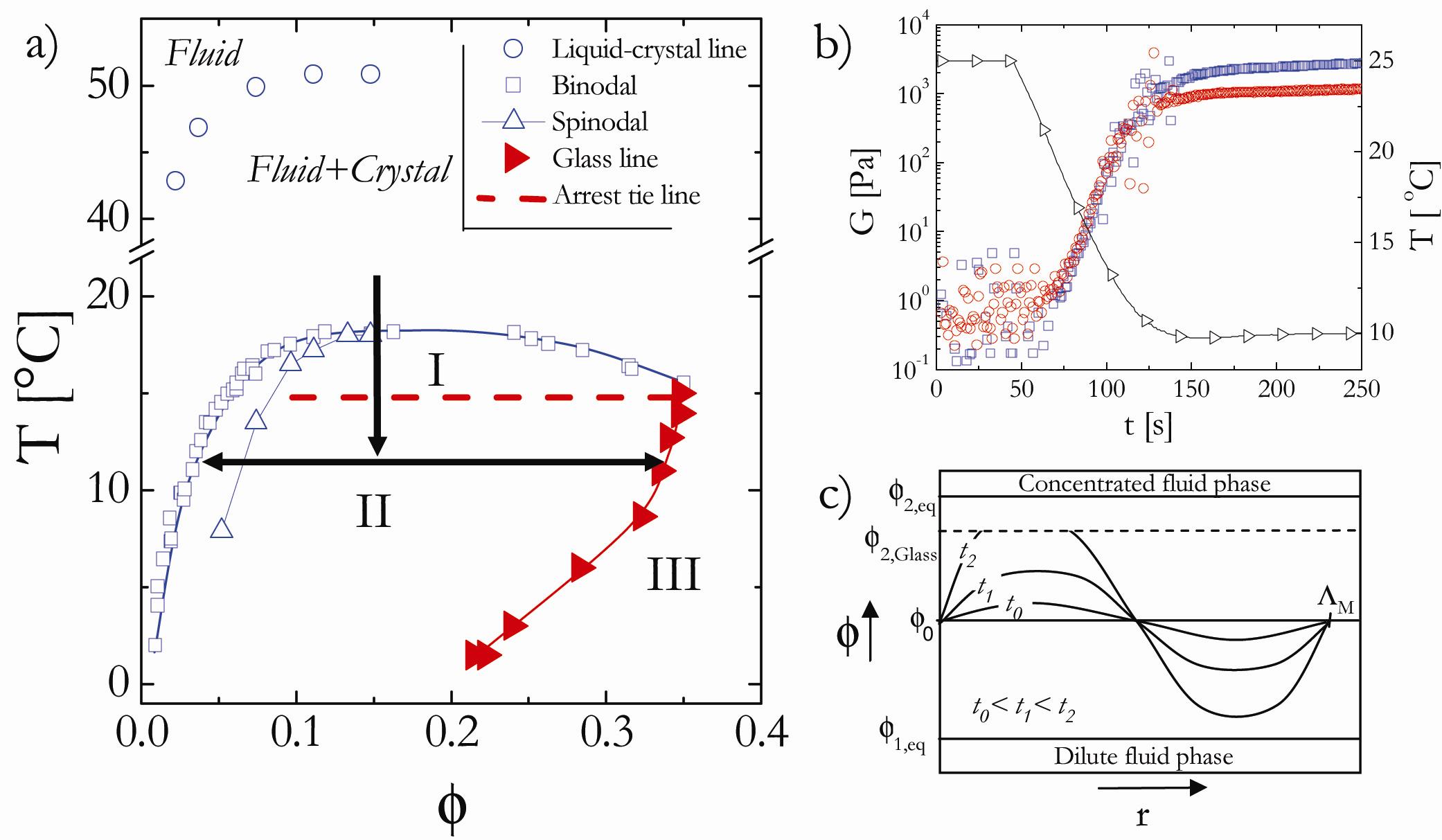}
 \caption{(a) State diagram of aqueous lysozyme solutions showing
 regions of crystal formation (open circles: experimentally determined phase boundary), and the metastable liquid-liquid phase separation below the experimentally deterimined coexistence curve (open squares). The region below the coexistence curve can be separated into three areas differing in their kinetic behavior: A region of complete demixing (I), gel formation through an arrested spinodal decomposition (II), and a homogeneous attractive glass (III). Full symbols stand for the results of the centrifugation experiments described in  \cite{2007PRL.cardinaux}. (b) Temporal evolution of the sample temperature and the concomitant values of the storage and loss moduli G' and G", respectively at a frequency of xx Hz during a temperature quench as indicated by the arrow in Fig. 1 a. (c) Schematic description of the model used to understand the behavior observed in Fig. 1 a and b. It schematically describes the evolution of
 density fluctuations  as a function of time ($t_{1}<t_{2}<t_{3}$) during the early stage of spinodal decomposition. The initial and equilibrium volume fractions of the dense
 and dilute phases are denoted $\phi_{0}$, $\phi_{2}$ and $\phi_{1}$.
 $\phi_{2,eq}$ and $\phi_{1,eq}$ are the volume fraction
 of the right and left branch of the coexistence curve. $\Lambda$ is the wavelength of the density fluctuations.
 At a certain moment $t_{3}$ the volume fraction $\phi_{2}$ of the dense phase
 intersects the dynamical arrest threshold $\phi_{2,Glass}$
 upon which phase separation gets pinned
 into a space spanning gel network with a characteristic length $\Lambda$}\label{Fig1}
  \end{center}
\end{figure}

The presence of an arrested spinodal decomposition has for example been demonstrated for concentrated solutions of the model protein lysozyme, which is known to exhibit a phase diagram that closely matches the predictions for colloids with short-ranged attractions \cite{2007PRL.cardinaux}. Here the mechanism underlying the liquid-solid transition encountered when quenching a protein sample deeply below the coexistence curve for liquid-liquid phase separation has been interpreted based on a state diagram shown in Fig.\ref{Fig1}. Here the arrest or gelation line extends deep into the unstable region, and arrest occurs as the spinodal decomposition process leads to a bi-continuous structure which gets "pinned" into a rigid self-supporting network when the concentration of the high density regions crosses the arrest line and subsequently undergo dynamical arrest. The findings described by Cardinaux et al. \cite{2007PRL.cardinaux} are however in sharp contrast to one of the central conclusions drawn in ref. \cite{2008nat.lu}, where a universal phase diagram for colloids with short-range attraction had been constructed that suggested that the gelation line coincides with the phase separation boundary in the Baxter model. These authors then concluded that the origin of dynamic arrest came from the dense phase undergoing an attractive glass transition at $\phi_g\approx0.55$, and claimed that the attractive glass line would thus not extend into the phase separation region but instead follow its high density boundary.

It is clear that this apparent discrepancy warrants further investigations. Moreover, given the fact that the location of the glass line in the experiments with lysozyme had been determined indirectly by exploiting the unusual rheological properties of the arrested samples, these findings need further support. We thus now aim at investigating the sequence of events that finally lead to an arrested spinodal decomposition in concentrated solutions of lysozyme, and at obtaining an improved understanding of the link between quench depth, initial concentration and the resulting structural and mechanical properties. The main focus of the present paper is on the determination the characteristic length scale of the structures  formed in the arrested spinodal decomposition as a function of  $\phi$ and final quench temperature $T_f$, and the attempt to explain the results based on known universal scaling behavior of spinodal decomposition combined with an arrest scenario based on the phase diagram postulated by Cardinaux et al. and reproduced again in Fig. \ref{Fig1}. In the first part we present the structure of lysozyme samples quenched into the region between the spinodal line and the tie line at 15$^{\circ}$C. We demonstrate that these samples exhibit indeed the classical features of spinodal decomposition. We are able to quantitatively follow the temporal evolution of the transient bicountinous network during the so-called intermediate regime, determine the characteristic length $\xi$ of the network from a combination of video microscopy and ultra-small angle light scattering, and verify that it follows the expected universal scaling relationship where $\xi$ evolves with time $t$ as t$^{1/3}$.

In the second part we then explore the structure of samples quenched in the spinodal region below the arrest tie line at 15$^{\circ}$C. While we also observe the initial formation of a bicountinous network, the structure becomes rapidly time-independent. We show that $\xi$ now depends mainly on the quench rate and the quench depth, but not on the volume fraction of lysozyme, and that the data are consistent with linear Cahn-Hillard theory, thus supporting a mechanism where the sample arrest in the early stage of spinodal decomposition consistent with the position of the dynamical arrest line as determined previously \cite{2007PRL.cardinaux}.

\section{Material and Methods}
We use hen egg white lysozyme (Fluka, L7651) dispersed in an aqueous
buffer (20 mM Hepes) containing $0.5$ M sodium chloride. Details of the sample preparation procedures can be found elsewhere \cite{2006JPCB.Stradner, 2007PRL.cardinaux}. Initially a suspension at $\phi\approx0.22$ is prepared in pure buffer without added salt, and its pH is adjusted to $7.8 \pm0.1$
with sodium hydroxide \cite{2006JPCB.Stradner, 2007PRL.cardinaux}.
We then dilute it with a NaCl-containing buffer at
pH$=7.8$ to a final NaCl concentration of $0.5$ M. Particular care is taken to avoid partial phase separation upon mixing by pre-heating
both buffer and stock solution well above the coexistence curve for
liquid-liquid phase separation (cf. Fig. 1). This procedure results in completely
transparent samples at room temperature with volume fractions
ranging from $\phi=0.01$ to $0.19$, where $\phi$ is obtained from
the protein concentrations $c$ measured by UV absorption
spectroscopy using $\phi = c/\rho$, where $\rho$ = 1.351 $g/cm^3$ is
the protein partial specific density, \cite{1976jbc.millero}.

We use a combination of ultra small angle light scattering (USALS) and video microscopy
to study the temporal evolution of the bicontinuous network structure during spinodal decomposition of the lysozyme solutions. We investigate two types of temperature quenches differing only
by their cooling rate. For both types of quenches,
a fresh
lysozyme dispersion is filled in a cuvette at a temperature around 25$^{\circ}$C, and the sample is then quenched below the spinodal line
to a final temperature $T_{f}$. We use either rectangular scattering cells with a short path length between 10 $\mu$m and 20 $\mu$m
to avoid multiple scattering for the USAL or capillary tubes (width 1 mm, depth 50 $\mu$m) for video microscopy. In the 'fast' quench the sample is pre-quenched for a minute in an ethanol bath at $T_{f}$. The sample is then
quickly transferred to the thermostated cuvette holder of
the experiment, also at $T_{f}$ . This procedure allows us to obtain the fastest experimental quenches possible with the
cell size required by the experimental set-up. We estimate the time it takes for the sample to go from
25$^{\circ}$C to $T_{f}$ to be less than 30 s. 'Slow' quenches result
from a simple cooling of the sample from 25$^{\circ}$C to
$T_{f}$ as the sample is inserted in the pre-thermostated sample holder of the experiment. It then takes about 100 s
to thermalize the sample to $T_{f}$.

The USALS setup covers
a range of scattering vectors 0.1$\leq$$q\leq$2 $\mu$m$^{-1}$. It is described in detail elsewhere \cite{2006jpcm.bhat}. Video microscopy is performed with a Leica DM-
IRB in phase contrast mode. To minimize heating during observation, we illuminate the sample only during the time required to take
pictures. Moreover we attenuate the heating effect of the
UV and IR part of the illumination spectrum by placing a beaker of water between the light source and the
sample. The microscope focus is set to image the mid-
plane of the sample to avoid the influence from wetting
effects at the interfaces between the lysozyme solution
and the capillary, and the objective aperture is opened to
its maximum diameter to minimize the depth of the focal
plan to a few $\mu$m. We carefully monitor the sample for crystallization and  disregard all samples where crystals form during the course of the experiment.

\section{'Classical' phase separation}
For shallow quenches below the spinodal curve but above the arrest
tie line at 15$^{\circ}$C (Fig. \ref{Fig1}), phase separation proceeds
classically via spinodal decomposition. After some time, we obtain
two homogeneous phases
separated by a sharp interface, where $\phi_{1,eq}$ and $\phi_{2,eq}$ are the volume fractions of the upper and lower phase, respectively, as given by left and right branches of the binodal curve in Fig.  \ref{Fig1}. We proceed with a 'slow' quench in this region and observe the spinodal
decomposition under the microscope. At early times, the micrographs
in Fig. 2a show the typical evolution of a bicontinuous network with
a characteristic size which increases with time, indicating a
coarsening of the structures formed during the spinodal
decomposition. After about $t\sim$100 s, the dense phase starts to
sediment due to the density mismatch. Micrographs indeed show an
increasingly blurred appearance when the focus is moved to the
upper part of the capillary. From then on the focus is maintained in
the lower part of the capillary tube. Finally, around $t$=1000 s, the
dense phase wets the bottom of the capillary and spreads to totaly
cover the bottom of the capillary tube.

\begin{figure}[]
\begin{center}
\includegraphics[width=240pt]{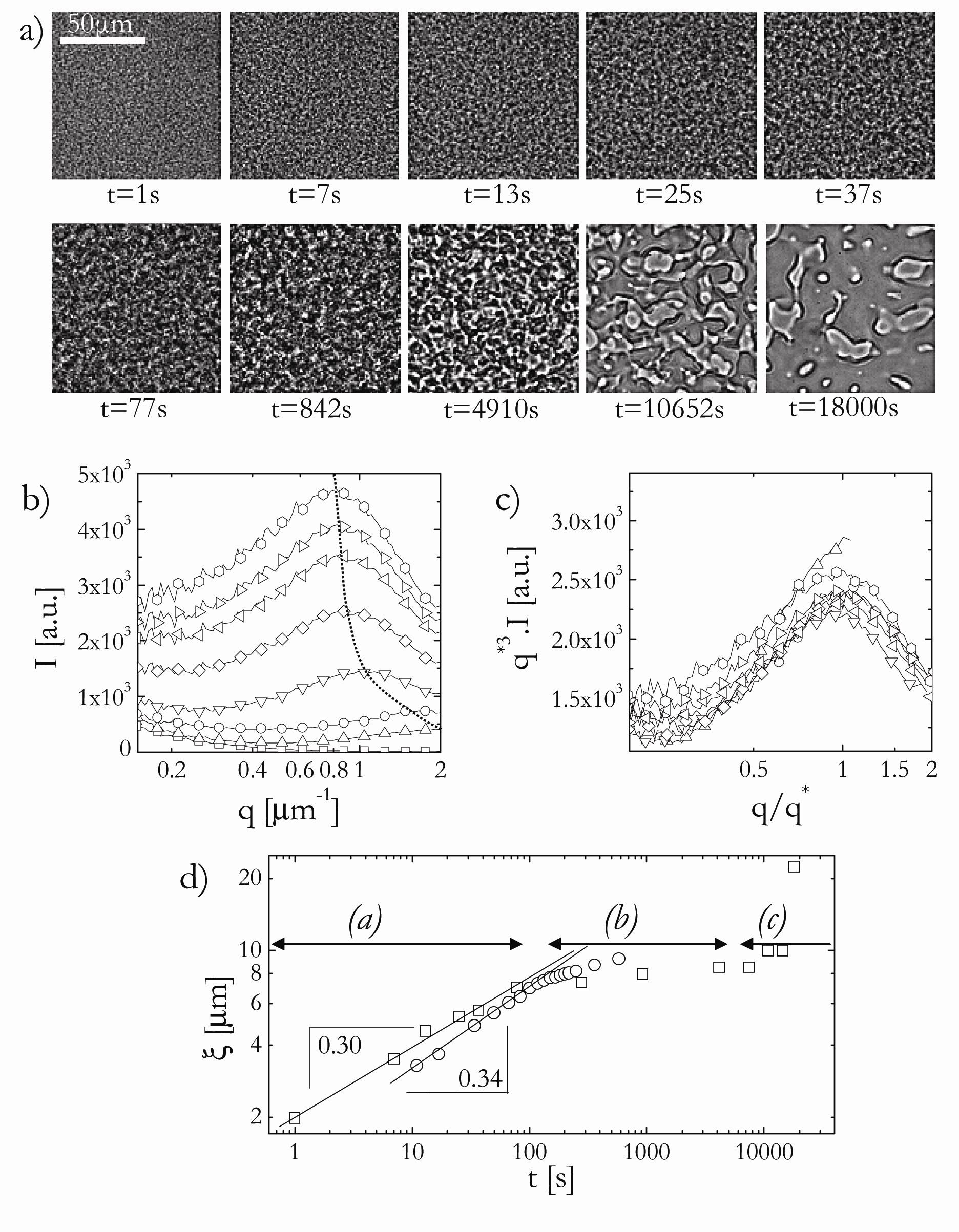}
 \caption{ Microscopy and USALS results obtained during 'classical' spinodal decomposition.
 (a) Micrographs showing the time evolution
 of the spinodal decomposition for a slow shallow quench at 17$^{\circ}$C and $\phi_{0}$=0.15.
  (b) USALS intensity as a function of $q$ showing the time evolution of the spinodal decomposition for a slow shallow quench at 16$^{\circ}$C and $\phi_{0}$=0.11. With increasing time the peak moves towards lower $q$-values: 0s ($\Box$), 11s ($\vartriangle$), 17s ($\circ$), 34s ($\triangledown$), 101s ($\diamond$), 118s ($\triangleleft$), 151s ($\triangleright$), 250s ($\circ$).
  (c) Dynamic scaling  $I(q)q^{*^{3}}$ of the data shown in (b).
  (d) Evolution of the characteristic length obtained during the microscopy and the USALS experiment for the two quenches shown in (a) and (c).}\label{Fig2}
  \end{center}
\end{figure}

In parallel USALS experiments were carried out. The evolution of the spinodal
decomposition process and the corresponding characteristic length
scale captured in Fourier space is shown in Fig. 2b. A peak is
observed in the scattering pattern at a scattering vector $q^{*}$
that corresponds to a characteristic length, $\xi=2\pi/q^{*}$. With
time the peak increases in amplitude and moves towards lower
scattering vectors, reflecting the coarsening of the structures formed
during the spinodal decomposition in regime \emph{(a)} already seen in microscopy. At $t\sim$100 s, the evolution is
perturbed either by sedimentation of the dense phase or by the
finite size of the cell.

Commonly one distinguishes between at least three
characteristic regimes during the spinodal decomposition process
\cite{1996.Debenedetti}: (\emph{i}) early, (\emph{ii}) intermediate (diffusibe) and (\emph{iii}) late (flow) stage. The early stage (\emph{i}) is described by linear Cahn-Hillard theory \cite{1958jcp.cahn, 1959jcp.cahn, 1961jcp.cahn} and in essence predicts an
increase of the amplitude at constant $q^{*}$ until the peak
spatial concentration fluctuations have reached the final coexisting
concentrations and domain growth and coarsening sets in. This regime is too fast and the structural length scale too small to be captured by USALS. During the intermediate and late stage,  coarsening is expected to occur according to a  power law $\xi\sim t^{\alpha}$, where the scaling exponent is $ \alpha=1/3$ in the intermediate (\emph{ii}) and $ \alpha= 1$ in the late
(\emph{iii}) stage of the coarsening regime \cite{1995macro.tromp,
1979pra.siggia}. A detailed calculation for the intermediate stage
including hydrodynamics reveals that the exponent $\alpha$ is
between 0.2 and 1.1, depending upon the relative importance of
hydrodynamic interactions \cite{1996jcp.dhont}.

The scattering patterns shown in Fig.2b are indeed
typical for spinodal decomposition. This is illustrated in
Fig.2c, where the data are shown to follow universal scaling by $I(q)q^{*^{3}}$ as proposed by
Furukawa \cite{1995physA.Furukawa} and derived by Dhont for the
spinodal decomposition of colloids in the initial and intermediate
stage including hydrodynamic interactions \cite{1996jcp.dhont}.

A quantitative temporal evolution of $\xi$ as obtained from the data shown in Fig. 2b through $\xi=2\pi/q^{*}$ is presented in Fig. 2d. It reflects the three
regimes previously described: \emph{(a)} growth ($t \lesssim$ 100 s),
\emph{(b)} sedimentation (100 s$ \lesssim t \lesssim $10000 s), and
\emph{(c)} wetting-spreading ($t \gtrsim $10000 s).
A comparison with the data from the microscopy experiments obtained by calculating the power spectrum of the micrographs is also given in Fig. 2d. We see that in the intermediate regime, and before the onset of perturbations such as sedimentation and finite cell size
effects, the scattering follows a power law of the form $\xi\sim t^{\alpha}$, with $ \alpha\approx1/3$, in agreement  with theoretical predictions
and other experiments with colloidal model systems. (see \cite{1996jcp.Verhaegh} and references
therein).

\section{Arrested spinodal decomposition}
Having verified the that the system exhibits all the characteristics of classical bicontinuous spinodal decomposition for quenches with $T_{f}\geq15^{\circ}$, we now follow the structural evolution in a sample that is quenched
into the spinodal region to a final temperature below the arrest tie
line at 15$^{\circ}$C. Here, the sample not only initially undergoes spinodal
decomposition into a bicontinuous network with protein-rich and
protein-poor domains, it also exhibits a non-ergodic liquid-solid transition as shown in Fig.1b \cite{2007PRL.cardinaux}. As these processes turn out to be too fast to be monitored by video microscopy and USALS, we focus on the final and time-independent structural properties of the arrested samples as a function of the parameters  initial volume fraction $\phi_{0}$, quench depth $T_{f}$ and quench rate.

\subsection{Influence of $\phi_{0}$}
We first look at the evolution of the resulting structure in samples undergoing an arrested spinodal decompositions as a function of the initial concentration $\phi_{0}$ at constant quench temperature $T_{f}$=10$^{\circ}$C. Fig.3a presents examples of micrographs obtained during 'fast' quenches at four different $\phi_{0}$. While the initial spinodal decomposition and subsequent arrest is too fast to be resolved and the data collected is time independent, the resulting structures and scattering patterns are very similar to the data obtained during 'classical' phase separation. We thus proceed with the same analysis to extract $\xi$ from microscopy and USALS, and the results from these measurements are summarized in Fig. 3b. Error bars are estimated based on 10 successive identical experiments with fresh lysozyme samples, and we find$\xi$ to vary at most within 10$\%$ of its mean value. As shown in Fig.3d, the characteristic length depends on the type of quenches, but for similar quench rate we observe no variation of $\xi$ with $\phi_{0}$.

\begin{figure}[]
\begin{center}
\includegraphics[width=240pt]{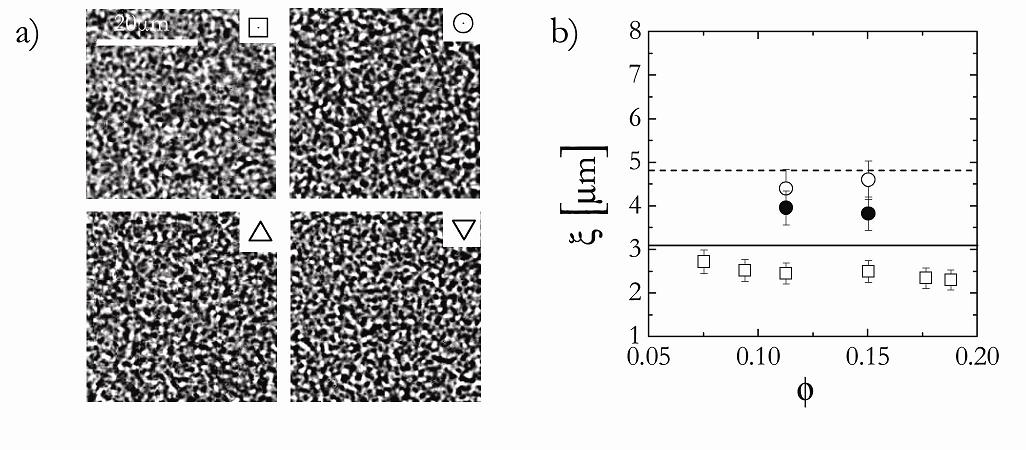}
 \caption{Microscopy and USALS results obtained during arrested spinodal decomposition at constant quench depth $T_{f}$=10$^{\circ}$C and for different $\phi_{0}$.
 (a) Micrographs showing arrested spinodal decomposition for a fast quench at:$\phi_{0}$=0.19 ($\Box$),$\phi_{0}$=0.15 ($\circ$), $\phi_{0}$=0.11 ($\vartriangle$), $\phi_{0}$=0.075 ($\triangledown$).
 (b) Evolution of the characteristic length as a function of $\phi_{0}$.
 ($\circ$) Results for a slow quench obtained during  microscopy experiments.
 ($\bullet$) Results for a slow quench obtained during  USALS experiments.
 ($\Box$) Results for a fast quench obtained during microscopy experiments.
}\label{Fig3}
  \end{center}
\end{figure}

The arrest scenario described in \cite{2007PRL.cardinaux} as sketched in Fig. 1c provides an explanation for these findings. In the early or linear stage, spinodal decomposition is triggered by density fluctuations of wavelength $\Lambda$ around $\phi_{0}$ \cite{1961jcp.cahn, 1996.Debenedetti, 1985.Laughlin, 1978jcp.mruzik}. The amplitude of the fluctuations increases with time until the low and high density values have reached their equilibrium values given by the corresponding points on the coexistence curve, i.e., $\phi_{1}$=$\phi_{1,eq}$ and $\phi_{2}$=$\phi_{2,eq}$. At this point the early stage of the spinodal decomposition ends and the structure starts coarsening. However, the results obtained by  \cite{2007PRL.cardinaux} indicate that in the arrested spinodal decomposition observed for lysozyme $\phi_{1}$ and $\phi_{2}$ never reach their equilibrium values because the dense phase becomes arrested in a glassy state at $\phi_{2}$=$\phi_{2,glass}<\phi_{2,eq}$. Therefore the characteristic length we observe corresponds to the the wavelength of the density fluctuations. The density fluctuation modes grow as $exp(Rt)$ where $R$ is the amplification factor. $R$ is proportional to the sum of two antagonistic terms: the second derivative of the Helmhotlz energy density of the bulk phase $\frac{\partial^{2}a}{\partial \phi^{2}}$ (which is also proportional to the isothermal compressibility) and the energy cost of a concentration gradient $\kappa$: $R\sim\frac{\partial^{2}a}{\partial \phi^{2}}-\kappa q^2$. The first term favors short wavelengths and it decreases as one approaches the spinodal line where it is equal to zero. The second term favors long wavelengths to minimize the number of interfaces in the system. The sharpness of the maximum of $R$ and the fact that $R$ enters the problem in an exponential function imply that the only density fluctuation that is relevant is the one for which $\Lambda$ maximizes $R$. Therefore, during the early stage of spinodal decomposition it is the fastest growing density fluctuation mode of wavelength $\Lambda_{M}\sim\sqrt{\kappa/ - \frac{\partial^{2}a}{\partial \phi^{2}}}$ that sets the characteristic length of the structure: $\xi$=$\Lambda_{M}(T_{f})$. In the $\phi_{0}$-series,  $\xi$ does not vary with $\phi_{0}$. The likely explanation is that $R$ and thus the derivative of the chemical potential  and the energy cost of a concentration gradient are constant. For those values to be constant the local volume fractions $\phi_{1}$ and $\phi_{2}$ must also be constant. This is the case in our experiments as it is only the total volume of each phase that varies with $\phi_{0}$: the total volume of the dense phase increases and the total volume of the dilute phase decreases with $\phi_{0}$.

\begin{figure}[]
\begin{center}
\includegraphics[width=240pt]{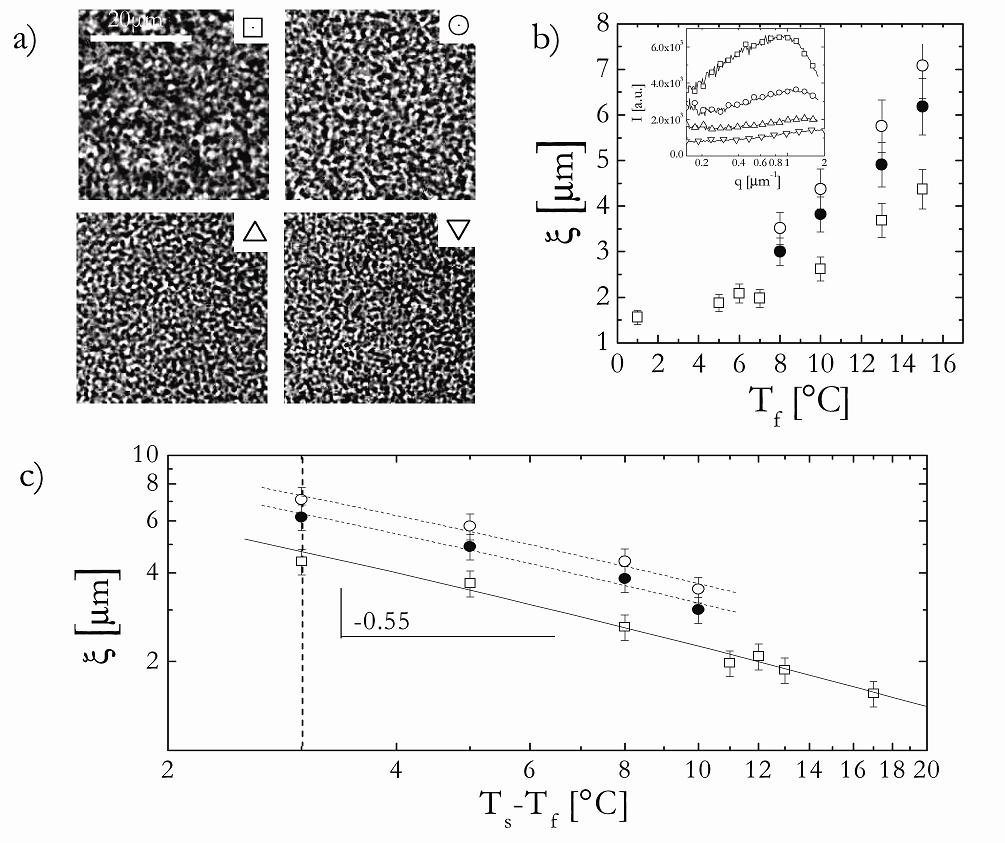}
 \caption{Microscopy and USALS results obtained during arrested spinodal decomposition at constant $\phi_{0}$=0.15 and for different quench depths $T_{f}$.
 (a) Micrographs showing arrested spinodal decomposition for a fast quench at: $T_{f}$=15$^{\circ}$C ($\Box$), $T_{f}$=10$^{\circ}$C ($\circ$), $T_{f}$=5$^{\circ}$C ($\vartriangle$), $T_{f}$=1$^{\circ}$C ($\triangledown$).
 (b) Evolution of $\xi$ as a function of $T_{f}$. ($\circ$) Results for a slow quench obtained during  microscopy experiments. ($\bullet$) Results for a slow quench obtained during USALS experiments. ($\Box$) Results for a fast quench obtained during microscopy experiments. The inset shows the evolution of the scattered intensity $I(q)$ with temperature for the four arrested samples measured with USALS, where the temperature increases from bottom to top.
 (c) Evolution of the characteristic length as a function of $T_{s}-T_{f}$ where $T_{s}$=18$^{\circ}$C is the temperature of the spinodal at $\phi_{0}$=0.15. Lines show linear fit of the data with a slope of 0.6$\pm$0.1 for ($\Box$), 0.55$\pm$0.1 for ($\bullet$) and 0.5$\pm$0.1 for ($\circ$). The dashed vertical line shows the position of the arrest tie line at $T_{f}$=15$^{\circ}$C.
}\label{Fig4}
  \end{center}
\end{figure}
\subsection{Influence of $T_{f}$}
A completely different behavior is observed when we look at the evolution of  $\xi$ for arrested spinodal decomposition at a constant volume fraction, $\phi_{0}$=0.15 as a function of $T_{f}$. In Fig.4a and b we show the micrographs and the corresponding  $\xi$ obtained from a 'fast' quench at four different $T_{f}$. In Fig.4b, we also show the evolution of the  $\xi$ as measured by video microscopy and USALS at four different $T_{f}$ for slow quenches. $\xi$ obviously depends on the quench rate, but for similar quench rates we observe a monotonic and pronounced increase of $\xi$ with decreasing quench depth, i.e., increasing $T_{f}$.

While $\xi$ is independent of  $\phi_{0}$ for fixed quench depth, Fig. 4b now clearly demonstrates that it strongly depends on $T_{f}$. We find that $\xi$  follows a critical scaling of the form $\xi\sim(T_{s}-T_{f})^{0.55}$ (Fig.4c) where $T_{s}$ is the temperature of the spinodal line. This finding further supports our assumption that arrest occurs during the early stage of spinodal decomposition. Indeed during the early stage of spinodal decomposition the wavelength $\Lambda_{M}$ of a density fluctuation is proportional to the square root of the energy cost of a concentration gradient divided by the second derivative of the Helmhotlz energy density of the bulk phase. $\Lambda_{M}$ diverges at the spinodal temperature because $\frac{\partial^{2}a}{\partial \phi^{2}}$ is equal to zero at $T_{s}$. The increase of $\xi$  with $T_{f}$ is qualitatively consistent with the fact that $\Lambda_{M}$ diverges at the spinodal temperature $T_{s}(\phi=0.15)$=18$^{\circ}$C. Around the critical volume fraction $\phi_{c}$, $\Lambda_{M}$ follows a critical scaling $\xi\sim(T_{s}-T_{f})^{\nu}$. Although $\phi_{0}$=0.15 is slightly below the critical volume fraction $\phi_{c}\simeq0.17$, the exponent found in our experiments $\nu=0.55\pm0.1$ is consistent with values of the critical exponent $\nu$ derived from  the mean field approximation ($\nu=0.5$) and from the renormalization theory ($\nu=0.63$) \cite{1985.Laughlin,1996.Debenedetti, 1986arpc.Sengers}.

The arrest scenario where the dense phase that forms during the early linear stage in the spinodal decomposition arrest as its concentration crosses the arrest line is further supported by the temperature dependence of $I(q)$ (Fig. 4b inset).  We observe that the amplitude of the peak decreases with decreasing temperature. This implies that the scattering contrast between the two phases is lower. This suggests that the difference between the two volume fractions decreases with temperature, which is again in agreement with the state diagram in Fig. 1a where the arrest line extends into the unstable region and moves to lower volume fractions with decreasing temperatures.

\subsection{Influence of the quench rate}
The quench rate influences the value of the characteristic length, with a fast quench leading to smaller $\xi$ as shown in Fig.3b and 4b. When quenches are slower, the sample spends more time at temperatures above $T_{f}$. Therefore density fluctuation modes that correspond to these temperatures  are activated and contribute to longer wavelengths. The combined effects of those different fluctuation modes results in a larger value of the resulting $\xi$.

\section{Conclusion}
We have previously shown in ref. \cite{2007PRL.cardinaux} that lysozyme solutions can undergo an arrested spinodal decomposition that results in the formation of solid-like samples. The rheological properties of these gels were found to exhibit an unusual frequency dependence with two well defined plateaus and associated yield stress values. This allowed us then to use centrifugation experiments to determine the local densities of both phases and to precisely locate the arrest line close to and within the unstable region of the phase diagram. Based on the extension of the arrest line in the coexistence region, we assumed that arrested spinodal decomposition occurs during the early stage when local variations of the density proceed. Our measurements now confirm this hypothesis. Using a combination of video microscopy and USALS we were able to demonstrate the formation of an arrested bicontinuous network for quenches below the arrest tie line at $T_{f}$=15$^{\circ}$C. Moreover, we have been able to show that the correlation length in these samples exhibits a temperature dependence that closely follows critical scaling expected  for density fluctuations during the early stages of spinodal decomposition.

These findings of course raise new questions. There is the apparent inconsistency with the study of Lu et al. \cite{2008nat.lu, 2008jpcm.Zaccarelli} where the authors had proposed that gelation for particles with short ranged attraction would happen through a spinodal decomposition that proceeds until the dense phase undergoes an attractive glass transition at $\phi_g\approx0.55$. One explanation could of course come from the fact that while lysozyme exhibits many of the features characteristic for colloids with short range attractions, it is also known that due to a slightly non-spherical shape and patchyness there are some non-centrosymmetric contributions to the interaction potential that one needs to account for in order to quantitatively reproduce the entire phase diagram  \cite{2008jcp.gogelin}. However, an alternative explanation could come from the way the concentration in the arrested phase has been determined by Lu et al. The authors used a confocal microscope and
determined the acceptance probability of a test particle randomly inserted into
their 3D images. This procedure is not unambiguous for highly ramified systems such as an arrested phase formed by attractive particles within a bi-continuous texture as seen for lysozyme. Under these conditions the method used by Lu et al. would not distinguish between space inside and outside the arrested phase, but yield the local density of individual gel strands within this phase rather then the  average concentration in the mesoscopic phase. In this context it is also important to note that even for the small protein lysozyme with a size of about 2 nm the extension of the bi-continuous texture as given by the correlation length is of the order of a few $\mu$m, i.e. about a thousand particles in diameter, and thus much larger than the volume in which the (local) concentration is determined when using a confocal microscope. It is clear that similar measurements as we have done for lysozyme would allow to help clarify this issue.


However, irrespective of the general validity of our findings with respect to the phase behavior of colloids, our results suggest some interesting possibilities. Given the strong interest in obtaining gel-like systems with tunable structural and viscoelastic properties for numerous applications in areas as diverse as materials and food science, arrested spinodal decomposition where the solidification occurs in the early stage offers means to create gels with a full control over the resulting correlation length of the bi-continuous network. This work also supports and gives new perspectives to the recent work of E. Dufresne where they have hypothesize that multiple lineages of birds have convergently evolved to exploit phase separation and kinetic arrest to self-assemble spongy color-producing nano-structures in feather barbs \cite{2009sm.dufresne}. Moreover, as we already know from our previous rheological measurements that the arrested networks have a material response function that is characterized by two distinct elastic moduli that we believe to be coupled to the two characteristic length scales $\xi$ and protein diameter \cite{2007PRL.cardinaux}, the obtainable variability in $\xi$ may result in large variations of the viscoelastic properties of these soft solids, and thus offer interesting possibilities for creating protein gels with tailored texture. We thus currently work on a full characterization of the viscoelastic properties of systems that result from different quenches and thus different structural properties.

\acknowledgments We are deeply grateful for fruitful discussions
with Veronique Trappe and Roberto Cerbino. This work
was supported by the Adolphe Merkle Foundation, the Swiss National Science Foundation, the State
Secretariat for Education and Research (SER) of Switzerland and the
Marie Curie Network on Dynamical Arrest of Soft Matter and Colloids
(MRTN-CT-2003-504712).

\include{BiblioAS}
\bibliography{bibliothese}

\end{document}